\documentclass[usenatbib]{mnras}
\usepackage{multirow}
\usepackage[utf8]{inputenc}
\usepackage[T1]{fontenc}
\usepackage{graphicx}	
\usepackage{amsmath}	
\usepackage{amssymb}	
\usepackage{multicol}   
\usepackage{bm}		    
\usepackage{pdflscape}
\usepackage{hyperref}
\usepackage{xcolor}

\title[The Sensitivity of the Redshift Distribution to Galaxy Demographics]{The Sensitivity of the Redshift Distribution to Galaxy Demographics}
\author[P. Sudek et al.]{
Philipp Sudek$^{1}$\thanks{E-mail: philipp.sudek@port.ac.uk},
Lucia F. de la Bella$^{1}$,
Adam Amara$^{1}$,
and William G. Hartley$^{2}$
\\
$^{1}$Institute of Cosmology and Gravitation, University of Portsmouth, Portsmouth, PO1 3FX, UK\\
$^{2}$Department of Astronomy, University of Geneva, ch. d’Ecogia 16, CH-1290 Versoix, Switzerland\\
}

\date{Accepted 2022 August 12. Received 2022 August 12; in original form 2022 January 11}

\pubyear{2022}

\begin{document}
\label{firstpage}
\pagerange{\pageref{firstpage}--\pageref{lastpage}}
\maketitle

\begin{abstract}
Photometric redshifts are commonly used to measure the distribution of galaxies in large surveys. However, the demands of ongoing and future large-scale cosmology surveys place very stringent limits on the redshift performance that are difficult to meet. A  new approach to meet this precision need is forward modelling, which is underpinned by realistic simulations. In the work presented here, we use simulations to study the sensitivity of redshift distributions to the underlying galaxy population demographics.  We do this by varying the redshift evolving parameters of the Schechter function for two galaxy populations, star-forming and quenched galaxies. Each population is characterised by eight parameters. We find that the redshift distribution of shallow surveys, such as SDSS, is mainly sensitive to the parameters for quenched galaxies. However, for deeper surveys such as DES and HSC, the star-forming parameters have a stronger impact on the redshift distribution. Specifically, the slope of the characteristic magnitude, $a_\mathrm{M}$, for star-forming galaxies has overall the strongest impact on the redshift distribution. Decreasing $a_\mathrm{M}$ by 148 per cent (its given uncertainty) shifts the mean redshift by ${\sim} 45$ per cent.
We explore which combination of colour and magnitude measurements are most sensitive to $a_\mathrm{M}$ and we find that each colour-magnitude pair studied is similarly affected by a modification of $a_\mathrm{M}$.
\end{abstract}


\section{Introduction}
The current standard model of cosmology, the $\Lambda$ cold dark matter ($\Lambda$CDM) model, describes a universe consisting of dark energy, cold dark matter and baryonic matter evolving with time. At the present time, the energy density of the Universe is mainly determined by the dark energy and the dark matter components.
However, these components are not fully understood yet and have to be better constrained \citep[e.g.][]{Clifton2012, Capela2013, Spergel2015}.  

With different cosmological probes, we can shed some light on the dark parts of cosmology and constrain the parameters of dark energy and dark matter. Type Ia supernovae, the cosmic microwave background (CMB) and weak gravitational lensing \citep[e.g.][]{Albrecht2006, Nicola2017} are important probes. For example, constraining the equation-of-state parameter is an important objective of cosmic shear surveys.
Most of these probes depend on the accurate prediction of the redshift distribution of galaxies, $n(z)$.
Using spectroscopy is the most precise way to determine the redshifts of galaxies.
By identifying absorption and emission lines, one can calculate the redshift of the corresponding object.
Although, these so-called `spectroscopic' redshifts are not a suitable choice for current and upcoming experiments such as the Dark Energy Survey\footnote{\url{https://www.darkenergysurvey.org}}; the Hyper Suprime-Cam\footnote{\url{https://hsc.mtk.nao.ac.jp/ssp/}} of the Subaru Telescope; the Sloan Digital Sky Survey\footnote{\url{https://www.sdss.org}}; the Rubin Observatory Legacy Survey of Space and Time\footnote{\url{https://www.lsst.org}} (LSST) and Euclid\footnote{\url{https://sci.esa.int/web/euclid/}}.
The big areas and depths observed by these surveys make performing the required spectroscopy extremely time-consuming and cost-intensive \citep{Amara2007, Abdalla2008}.

As an example, the Dark Energy Survey \citep[DES,][]{Abbott2005} is a ground-based telescope observing 5000\,deg$^2$ of the Southern Sky in the visible spectrum. 
It uses the Dark Energy Camera \citep[DECam,][]{Honscheid2008, Flaugher2015} to observe more than 300 million galaxies. 
The DES standard bandpasses are the DECam $g$, $r$, $i$, $z$ and $Y$ filters\footnote{\url{http://www.ctio.noao.edu/noao/node/13140}} spanning a wavelength range from 400 to 1065\,nm with a depth of $i = 23.44$ \citep[in Data Release 1,][]{Abbott2018}.

Another example is the Hyper Suprime-Cam of the Subaru Telescope\footnote{\url{https://subarutelescope.org/en/}} \citep[HSC,][]{Miyazaki2012}. It covers 1400\,deg$^2$ in the wide field, 27\,deg$^2$ in the deep field and 3.5\,deg$^2$ in the ultra-deep field.
HSC consists of five broad-band filters and four narrow-band filters \citep{Kawanomoto2018}. The broad-band filters, $g$, $r$, $i$, $z$ and $y$, cover a wavelength range of ${\sim}400$\,nm to ${\sim}1100$\,nm.
The $i$-band limiting magnitude is $26.2$ in the wide layer, $27.1$ in the deep layer and $27.7$ in the ultra-deep layer \citep{Aihara2018}.

The Sloan Digital Sky Survey \citep[SDSS,][]{York2000} has a lower depth. While taking data in the $u$, $g$, $r$, $i$ and $z$ filter bands\footnote{Note that although we used the same notation for some of the DES, HSC and SDSS filter bands, we mean the filters specific to the surveys. For example, the $g$-band of DES, HSC and SDSS are different.} the $i$-band limiting magnitude is $21.3$ \citep{Stoughton2002}.
The covered area is 14555\,deg$^2$. Therefore, SDSS is measuring a much larger part of the sky compared to DES and HSC.
The filters of SDSS cover a wavelength range of $300$\,nm to $1100$\,nm\footnote{\url{https://www.sdss.org/instruments/camera/}}.  

For such big surveys, photometry is an alternative to spectroscopy. Photometric redshifts can be determined in two ways. First, one can fit the spectral energy distribution (SED) in different broad-band filters. The approach is to use different SED templates and fit them to the observed fluxes. Through this process, one gains information about the galaxies' redshifts \citep{Bolzonella2000, Benitez2000}.
Alternatively, photometric redshifts can be estimated by using machine learning methods \citep{Kind2013, Sadeh2016}. This would require a sample of known redshifts to train the algorithm that determines the redshifts of the investigated sample.
With template fitting and machine learning methods, we are able to get redshift estimates even for deep surveys. The problem is, the constraining power of cosmological probes that require redshift distributions is decreased by the lack of precision in the photometric redshifts \citep{Bordoloi2010, Bordoloi2012, Salvato2019}.  

To solve the problems of spectroscopic and photometric redshifts, \cite{Herbel2017} developed a method to simulate the redshift distribution directly.
The authors used Approximate Bayesian Computation with a forward modelling approach to determine a full posterior of $n(z)$ without measuring the individual redshifts and which can be used in measurements.
Thus, one does not rely on the technically challenging spectroscopic redshifts for big surveys. Furthermore, this approach delivers a more precise estimate of the redshift distribution than photometry. 

In this paper, our main goal is to study the sensitivity of the redshift distribution of sets of galaxies to the underlying galaxy population properties. The motivation is to better understand which features of galaxy populations need to be well understood to make the next generation of precision cosmological measurements.

To achieve our goal, we simulate galaxy catalogues solely with \cite{Herbel2017}'s forward modelling approach based on redshift-dependent Schechter luminosity functions, which describe the populations of different galaxy types. We analyse how changes of the Schechter luminosity function impact the redshift distribution for different survey types. 
Precisely, we identify the model parameters (Schechter parameters) that affect the simulations of the redshift distribution for the different survey types the most.
Hence, this work demonstrates which parameters need better constraints to improve the precision of the simulated redshift distribution. This information will be important for the analysis of next generation precision measurements. 

Furthermore, we investigate how the Schechter parameters affect the galaxy observables (apparent magnitude and colour distributions). As the magnitude and colour distributions are directly measurable quantities, this helps us to understand how the model parameters impact observations. In addition to this, this analysis identifies the observables that are most sensitive to the Schechter parameters and could help to constrain these parameters in the future.

In our analysis, we considered two different galaxy populations, a star-forming and a quenched model.
Additionally, we chose the same parametric model as \cite{Herbel2017} to assign spectra to the galaxies. These spectra are used to calculate the apparent magnitudes of the galaxies.  
We simulated observed galaxy catalogues for DES-, SDSS- and HSC-like surveys. The catalogues contain the redshift values, absolute Johnson $B$-band magnitudes, intrinsic\footnote{Observational error not considered} apparent magnitudes in different filter bands, observed\footnote{The word "observed" indicates that the magnitudes include an observational error. We describe the simulation of the error in section \ref{sec:error-models}.} apparent magnitudes in different filter bands and the errors of the apparent magnitudes.

To perform the simulations, we used SkyPy \citep{SkyPyCollaboration2020, Amara2021}. SkyPy is a publicly available python package for simulating the astrophysical sky and includes physical and empirical models for different observables. Therefore, SkyPy can be used to perform end-to-end simulations of these observables.
We extended the code to account for observational errors and to obtain the observed magnitudes. A detailed description of this is summarised in section \ref{sec:setup}.

Furthermore, to account for signal-to-noise and saturation effects, we applied survey specific magnitude cuts in the $i$-band that remove too bright and too faint galaxies.
We investigated the sensitivity of the redshift, colour and magnitude distributions to the different Schechter parameters by repeating the simulation after changing one parameter value by the range allowed by its uncertainty. The resulting catalogue was used as our test catalogue and compared to the fiducial catalogue\footnote{Catalogue that was simulated with the default set of Schechter luminosity function parameters. See sections \ref{sec:setup} and \ref{sec:method}.}.

This paper is structured as follows. In section \ref{sec:model}, we summarise the mathematical models of our simulations. 
Section \ref{sec:setup} presents our simulation set-up including the fiducial Schechter parameters, a detailed description of our error model and how we perform the magnitude cuts.
In section \ref{sec:method}, we describe the methodology of our simulations and the comparison of fiducial and test catalogue.
Our results are presented in section \ref{sec:results} and section \ref{sec:conclusion} includes our conclusions.
Throughout this work, we use a standard $\Lambda$CDM cosmology with $h = 0.7$, $\Omega_\mathrm{m} = 0.3$ and $\Omega_\Lambda = 0.7$.

\section{Forward Modelling Galaxy Catalogues}
\label{sec:model}
In this section, we describe our model for simulating galaxy catalogues including redshifts, magnitudes and colours. Our simulations follow a forward modelling approach to draw galaxies from the Schechter luminosity function, as described by \cite{Herbel2017}.
We assign the redshifts and absolute $B$-band magnitudes that follow the Schechter luminosity function to our objects. With these quantities, we calculate the apparent magnitudes in different filter bands by simulating the spectra that are based on template coefficients.
A deeper investigation and motivation of this model will be the scope for future work.

\subsection{Luminosity Function and Galaxy Numbers}
Generating intrinsic galaxy catalogues requires a model of the expected galaxy number in a certain volume of the sky. The luminosity function $\Phi$ describes the number of galaxies $N$ per comoving volume $V$ and absolute magnitude $M$ as a function of redshift $z$,
\begin{equation}
\label{eq:luminosity_function_definition}
    \Phi(z,M) = \frac{\mathrm{d}N}{\mathrm{d}V\,\mathrm{d}M}\;.
\end{equation}
Different observations \citep{Loveday2012, Lopez-Sanjuan2017} have supported the work from \cite{Schechter1976} that described the functional form of the luminosity function.
The Schechter luminosity function is given as
\begin{equation}
\label{eq:schechter}
\begin{split}
    \Phi(z, M) =\; &0.4\, \ln(10)\, \Phi_*(z)\, 10^{0.4(M_*(z) - M)(\alpha(z)+1)} \\ 
    &\cdot \exp\left(-10^{0.4(M_*(z) - M)}\right)\:,
\end{split}
\end{equation}
where the faint-end-slope $\alpha$, the characteristic magnitude $M_*$ and the amplitude $\Phi_*$ are free-fitting parameters. These parameters depend on the type of the galaxy population and on the redshift. 

The evolution of the luminosity function with redshift can be empirically motivated \citep{Herbel2017}. 
Keeping $\alpha$ constant for each galaxy population, the characteristic magnitude and the amplitude are parametrised as
\begin{align}
M_*(z) &= a_\mathrm{M}\,z + b_\mathrm{M} \label{eq:Mstar}\,, \\
\Phi_*(z) &= b_\phi \exp(a_\phi z) \label{eq:Phistar}\;.
\end{align}
Combined each population is described by five parameters. However, we assume that each type evolves according to the same functional form.
In this paper, we refer to the parameters $a_\mathrm{M}$, $b_\mathrm{M}$, $a_\phi$ and $b_\phi$ as the ``Schechter parameters'' as these are the model parameters we investigated.

Considering the volume element as a light cone, we can write the comoving element in terms of the solid angle $\Omega$ and redshift $z$ as
\begin{equation}
\label{eq:comoving_element}
    \mathrm{d}V = \frac{d_\mathrm{H} d_\mathrm{M}^2}{E(z)} \mathrm{d}\Omega \, \mathrm{d}z\;,
\end{equation}
where $d_\mathrm{H}$ is the Hubble distance, $d_M$ is the transverse comoving distance and $E(z) = (\Omega_\mathrm{m}(1+z)^3 + \Omega_\mathrm{k}(1+z)^2 + \Omega_\Lambda)^{0.5}$.
Plugging this into equation \ref{eq:luminosity_function_definition}, we get the number of galaxies per absolute magnitude, redshift and solid angle
\begin{equation}
\label{eq:finalSchechter}
    \phi(z,M) = \frac{\mathrm{d}N}{\mathrm{d}M\,\mathrm{d}\Omega\,\mathrm{d}z} = \frac{d_\mathrm{H} d_\mathrm{M}^2}{E(z)} \Phi(z,M)\;.
\end{equation}

This function is the number density which can be used to draw samples of galaxies. These samples contain redshift values and absolute magnitudes that follow the distribution $\phi$.
Additionally, by integrating equation \ref{eq:finalSchechter}, one obtains the number of galaxies that would be observed in a light cone, spanned by solid angle $\Omega$ with redshifts between $z_1$ and $z_2$ and magnitudes between $M_1$ and $M_2$,
\begin{equation}
\label{eq:total_number_from_schechter}
    N(z_1,z_2,M_1,M_2,\Omega) = \Omega \int_{M_1}^{M_2} \int_{z_1}^{z_2} \phi(z,M)\, \mathrm{d}z\,\mathrm{d}M\;.
\end{equation}

\subsection{Galaxy Colours}
The colour of a galaxy is defined as the difference in its magnitudes in two different filter bands. Given the spectrum $f(\lambda)$ of a galaxy, calculating the magnitude in a particular band is straightforward.

First, the absolute magnitude in this band has to be calculated. In the AB magnitude system, this is given as \citep{Blanton2003}
\begin{equation}
\label{eq:absolute_magnitude}
    M_i = -2.41 - 2.5 \log_{10}\left[\frac{\int_0^\infty \mathrm{d}\lambda_\mathrm{o} \lambda f(\lambda) R(\lambda_\mathrm{o})}{\int_0^\infty \mathrm{d}\lambda_\mathrm{o} \lambda^{-1} R(\lambda_\mathrm{o})}\right]\,,
\end{equation}
where $R(\lambda)$ is the filter response giving the fraction of photons that are included in the signal. The subscript `o' indicates the observed frame and the observed and emitted wavelengths are related through the redshift relation $\lambda_o = (1+z) \lambda$. The units of the spectrum $f(\lambda)$ are erg\,s$^{-1}$\,cm$^{-2}$\,\AA$^{-1}$ in this equation. 
To transform this into an apparent magnitude, we
employ the distance modulus, $\mathrm{DM}(z) = 5\log_{10}(d_\mathrm{L}(z)/10\,\mathrm{pc})$, where $d_\mathrm{L}$ is the luminosity distance.
Thus, the apparent magnitude in band $i$ is given by
\begin{equation}
\label{eq:apparent_magnitude}
    m_i = M_j + \mathrm{DM}(z) + K_{ji}(z) \;,
\end{equation}
with $K_{ji}(z)$ accounting for the $K$ correction \citep[][]{Hogg2002,Blanton2007}, which is relating the emitted rest-frame magnitude in broad photometric bandpass j to the observed
apparent magnitude in band i.

\subsection{Galaxy Spectra}
Simulating the magnitudes of the galaxies requires the galaxies' spectra. To model the spectrum, we again follow the approach of \cite{Herbel2017}, which is based on the work by \cite{Blanton2007}. The flux density as it would be seen if the galaxy is at a distance of 10\,pc is given as a linear combination of five template spectra $t_i(\lambda)$,
\begin{equation}
\label{eq:sed}
    f(\lambda) = \sum_i c_i t_i(\lambda)\,,
\end{equation}
where the units of $f(\lambda)$ are erg\,s$^{-1}$\,cm$^{-2}$\,\AA$^{-1}$ per solar mass.
\cite{Herbel2017} showed that after re-weighting the coefficients $c_i$ by a weight $w_i$, the coefficients follow a Dirichlet distribution of order five so that $\sum_i \tilde{c}_i = 1 $ with $\tilde{c}_i = c_i/w_i$.
A Dirichlet distribution \citep[][]{Kotz2019} of order $K \ge 2$ is defined as

\begin{equation}
    \mathrm{Dir}(x_1, ..., x_K; \alpha_1, ..., \alpha_K) = \frac{1}{\mathrm{B}(\bm{\alpha})} \, \prod_{i=1}^K x_i^{\alpha_i - 1} \;,
\end{equation}
where $\bm{\alpha} = (\alpha_1, ..., \alpha_K)$ with $\alpha_i > 0$ are parameters characterising the Dirichlet distribution and $\mathrm{B}(\bm{\alpha})$ is the multivariate beta function. 

Therefore a Dirichlet distribution of order five is characterised by five parameters $\alpha_i$. These parameters are redshift dependent in general. Once again according to \cite{Herbel2017}, we assume that the evolution of the parameters is described as
\begin{equation}
\label{eq:dirichlet}
    \alpha_i(z) = (\alpha_{i,0})^{1-z/z_1} \cdot (\alpha_{i,1})^{z/z_1}\;.
\end{equation}
The two parameters $\alpha_{i,0}$ and $\alpha_{i,1}$ describe the galaxy population at redshift $z=0$ and $z=z_1>0$, respectively. They are also different for each galaxy type.

Note that the K-correct templates $t_i(\lambda)$ are intended to fit broad-band photometry, which is suitable for our case. In this work, we decided to build on the studies presented in \cite{Herbel2017}, \cite{Fagioli2018, Fagioli2020} and \cite{Tortorelli2018, Tortorelli2020, Tortorelli2021}, which show that the K-correct templates can be used in our analysis, especially at low redshifts.
We are also revisiting the quality of the templates for higher redshifts in a new study (Hartley et al. in prep), which will show that this model predicts colour and apparent magnitude distributions even for samples with higher redshifts.

\section{Simulation Setup}
\label{sec:setup}
In this section, we describe the setup of our simulations. We summarise how the fiducial catalogues of each survey were generated based on the model described in section \ref{sec:model}. 
That includes the Schechter parameters, the Dirichlet parameters for generating the spectra, the simulated sky area and filter bands of each survey type.

Furthermore, we describe how the intrinsic catalogues are transformed into observed galaxy catalogues for each survey. Additionally, we explain our choice of magnitude cuts to mock real observations.

\subsection{Schechter Parameters}
\label{sec:schechter_params}
In our work, we used the parameters from \cite{Tortorelli2021} following a model consisting of star-forming and quenched galaxies. Table \ref{tab:schechterParams} summarises the values of the Schechter parameters together with their errors.

As aforementioned, each galaxy population can be simulated by drawing from the same distribution with different parameters.
The redshift dependency of our model is given by the faint-end slope $\alpha$, the characteristic magnitude $M_*$ and the amplitude of the Schechter function $\Phi_*$.
In equations \ref{eq:Mstar} and \ref{eq:Phistar}, we parametrise $M_*$ as a linear function and $\Phi_*$ as an exponential function of redshift $z$.
Therefore, our model is described by five parameters, $\alpha$, $a_\mathrm{M}$, $b_\mathrm{M}$, $a_\phi$ and $b_\phi$. 

In our investigation, $\alpha$ was considered redshift independent such that only four parameters characterised the evolution with redshift.
We focused on this set of parameters and investigated the sensitivity of the redshift distribution to a change in these parameters for the different surveys.
Since the parameters have different values according to the galaxy type, we had one set of four parameters for the star-forming galaxies and another set of four parameters for the quenched galaxies.

\begin{table}
\centering
 \caption{Schechter Parameters for our simulations. Values and errors are taken from \protect\cite{Tortorelli2021}. The positive and negative errors correspond to the 84th percentile and 50th percentile values, respectively.}
 \label{tab:schechterParams}
 \begin{tabular}{ccc}
  \hline
  Parameter & Star-Forming & Quenched \\
  \hline
  $\alpha$ & $-1.3$ & $-0.5$ \\[2pt]
  $a_\mathrm{M}$ & $-0.439^{+0.535}_{-0.652}$ & $-0.697^{+0.698}_{-0.729}$ \\[3pt]
  $b_\mathrm{M}$ & $-20.623^{+0.417}_{-0.425}$ & $-20.372^{+0.513}_{-0.466}$ \\[3pt]
  $a_\phi$ & $-0.088^{+0.297}_{-0.277}$ & $-0.836^{+0.812}_{-0.733}$ \\[3pt]
  $b_\phi$ & $0.004245^{+0.001337}_{-0.001452}$ & $0.005169^{+0.003112}_{-0.003515}$ \\
  \hline
 \end{tabular}
\end{table}

\subsection{Dirichlet Parameters}
\label{sec:dirichlet_params}
Table \ref{tab:dirichletParams} summarises the Dirichlet parameters from \cite{Tortorelli2018} and the weights from \cite{Herbel2017}.

As we explained in section \ref{sec:model}, we simulated the spectra based on template coefficients that are drawn from a Dirichlet distribution. We considered a model independent of redshift such that $\alpha_i(z)$ from equation \ref{eq:dirichlet} is a constant (just denoted as $\alpha_i$ from here on).

To account for the different galaxy types, we also introduced weights for each coefficient (more details in \cite{Herbel2017}).
Through the distinction between star-forming and quenched galaxies, we had a total of ten parameters and their corresponding weights, respectively.

\begin{table}
\centering
 \caption{Dirichlet parameters $\alpha_i$ and weights $w_i$ for our simulations. The Dirichlet parameters are taken from \protect\cite{Tortorelli2018} and the weights are from \protect\cite{Herbel2017}.}
 \label{tab:dirichletParams}
 \begin{tabular}{ccc}
  \hline
  Parameter & Star-Forming & Quenched \\
  \hline
  $\alpha_1$ & $1.9946549$ & $1.62158197$ \\[3pt]
  $\alpha_2$ & $1.99469164$ & $1.62137391$ \\[3pt]
  $\alpha_3$ & $1.99461187$ & $1.62175061$ \\[3pt]
  $\alpha_4$ & $1.9946589$ & $1.62159144$ \\[3pt]
  $\alpha_5$ & $1.99463069$ & $1.62165971$ \\[3pt]
  $w_1$ & $3.47 \cdot 10^9$ & $3.84 \cdot 10^9$ \\[3pt]
  $w_2$ & $3.31 \cdot 10^6$ & $1.57 \cdot 10^6$ \\[3pt]
  $w_3$ & $2.13 \cdot 10^9$ & $3.91 \cdot 10^8$ \\[3pt]
  $w_4$ & $1.64 \cdot 10^{10}$ & $4.66 \cdot 10^{10}$ \\[3pt]
  $w_5$ & $1.01 \cdot 10^9$ & $3.03 \cdot 10^7$ \\
  \hline
 \end{tabular}
\end{table}

\subsection{Surveys and Observed Catalogues}
\label{sec:surveys}
We identified the Schechter parameters that have the biggest impact on the simulated redshift distribution of different surveys. Within the context of the \cite{Herbel2017} forward modelling approach, having more precise parameter values directly translates into better cosmological constraints that are based on the redshift distribution.

Moreover, we analysed the effect of the Schechter parameters on three survey types, based on their depths and fields of view. We investigated a DES-like survey, a deeper but narrower HSC deep-field-like survey and a shallower but wider SDSS-like survey.
These surveys have different sets of filter bands, which we needed to reflect in the simulations. 

With the SkyPy python package, we applied these filters directly. We chose to simulate the DES-like survey with the DECam $g$, $r$, $i$ and $z$ filters. 
In the case of HSC, we used the $g$, $r$, $i$, $z$ and $y$ filters.
To simulate the SDSS-like catalogue, we used the SDSS $u$, $g$, $r$, $i$ and $z$ filters.

Since the leading order effect of precision measurements is usually the number of galaxies, we simulated different fields of view for each survey. Thus, we had approximately the same number of galaxies in each fiducial catalogue and could explore the trends beyond pure number counts. We chose a simulated area of 10\,deg$^2$ for DES, 1.5\,deg$^2$ in the case of HSC and 135\,deg$^2$ for SDSS. 
That resulted in ${\sim} 58500$ galaxies in the DES-like fiducial observed catalogue, ${\sim} 59300$ in the SDSS-like catalogue and ${\sim} 58400$ galaxies in the HSC-like catalogue.
A summary of the filter bands, sky area and the resulting number of galaxies of our fiducial catalogues is given in table \ref{tab:summarySurveys}.

\begin{table}
\centering
 \caption{Summary of the investigated surveys including the simulated filter bands, sky area and the resulting number of galaxies of each fiducial catalogue.}
 \label{tab:summarySurveys}
 \begin{tabular}{cccc}
  \hline
  Survey & Filter Bands & Sky Area & Number of Galaxies\\
  \hline
  DES & $g$, $r$, $i$, $z$ & 10\,deg$^2$ & ${\sim} 58500$ \\[3pt]
  SDSS & $u$ ,$g$, $r$, $i$, $z$ & 135\,deg$^2$ & ${\sim} 59300$ \\[3pt]
  HSC & $g$, $r$, $i$, $z$, $y$ & 1.75\,deg$^2$ & ${\sim} 58400$ \\
  \hline
 \end{tabular}
\end{table}

To make the intrinsic catalogues of the surveys observed catalogues, we performed two additional steps in our simulations:

\begin{enumerate}
    \item Adding survey specific errors to the simulated intrinsic magnitudes
    \item Applying survey specific magnitude cuts 
\end{enumerate}

\subsubsection{Modelling Magnitude Uncertainties}
\label{sec:error-models}
In sections \ref{sec:schechter_params} and \ref{sec:dirichlet_params}, we described the setup to simulate intrinsic catalogues.
However, making realistic simulations of galaxy catalogues implies modelling the magnitude uncertainties appropriately.
These uncertainties must be specific to the survey. 
We used \cite{Rykoff2015}'s model to simulate the uncertainty, $\sigma_m$, of the apparent magnitudes in each filter band for every survey,

\begin{equation}
\label{eq:sigmaM}
    \sigma_m(F;F_\mathrm{noise}, t_\mathrm{eff}) = \frac{2.5}{\ln{10}} \left[ \frac{1}{Ft_\mathrm{eff}} \left( 1 + \frac{F_\mathrm{noise}}{F} \right) \right]^{1/2} \;,
\end{equation}
where 
\begin{equation}
\label{eq:flux}
    F=10^{-0.4(m - m_\mathrm{ZP})}   
\end{equation}
is the galaxy's flux,
\begin{equation}
\label{eq:fluxNoise}
    F_\mathrm{noise} = \frac{F_\mathrm{lim}^2 t_\mathrm{eff}}{10^2} - F_\mathrm{lim}
\end{equation}
is the effective noise flux and $t_\mathrm{eff}$ is the effective exposure time\footnote{Note that we absorbed the normalisation constant $k$, which is mentioned in \cite{Rykoff2015}, in the definition of $t_\mathrm{eff}$.}.
Furthermore, $m$ is the galaxy's magnitude, $m_\mathrm{ZP}$ is the zero-point magnitude of the filter band and $F_\mathrm{lim}$ is the $10\sigma$ limiting flux.

Equations \ref{eq:sigmaM}, \ref{eq:flux} and \ref{eq:fluxNoise} then define the uncertainty of the magnitude, $\sigma_m(m;m_\mathrm{lim}, t_\mathrm{eff})$, depending on magnitude $m$, limiting magnitude $m_\mathrm{lim}$ (magnitude associated with $F_\mathrm{lim}$) and effective exposure time $t_\mathrm{eff}$.
We also used the same model as \cite{Rykoff2015} for the effective exposure time,

\begin{equation}
\label{eq:exposureTime}
    \ln{t_\mathrm{eff}} = a + b(m_\mathrm{lim} - 21)\;,
\end{equation}
where $a$ and $b$ are free parameters.

We fitted $\sigma_m(m)$ (i.e. fitting $a$ and $b$) to measured data. In this way, we got the magnitude uncertainties specific to the survey. For the zero-point magnitudes $m_\mathrm{ZP}$, we used a value of 30 for each filter band and every survey.
That is justified because this model is only an approximation and has initially been developed for SDSS data only. The model is empirically motivated and does not include any deeper analysis of signal-to-noise effects that exist during observations. 

\begin{figure}
 \includegraphics[width=\columnwidth]{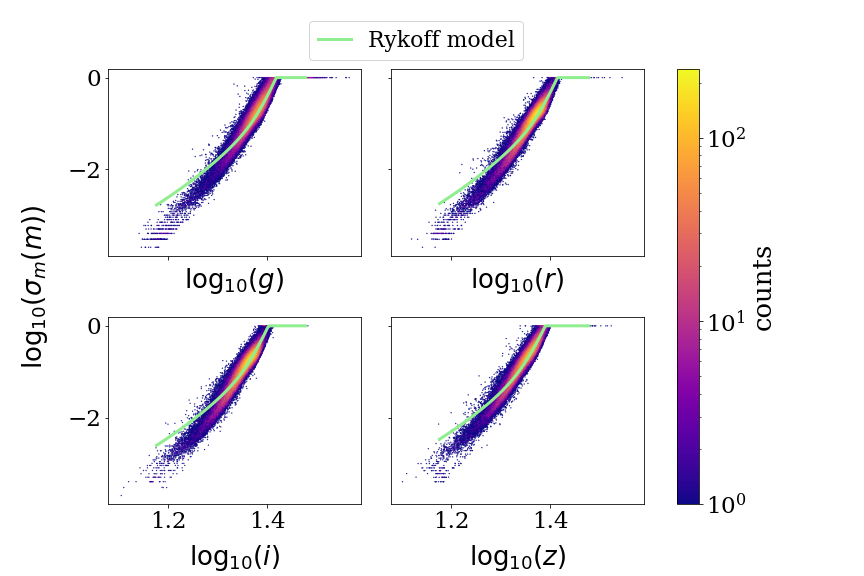}
 \caption{Magnitude uncertainties of a \textbf{DES-like survey}. The binned data points} show the DES Y1 measurements \citep{Drlica-Wagner2018} of the apparent magnitudes and its corresponding uncertainties, $\sigma_m(m)$, in the $g$, $r$, $i$ and $z$ filter bands. The green lines show the respective fits to the \protect\cite{Rykoff2015} error model. We see that model and data generally match.
 \label{fig:mUncertaintyDES}
\end{figure}

\begin{figure}
 \includegraphics[width=\columnwidth]{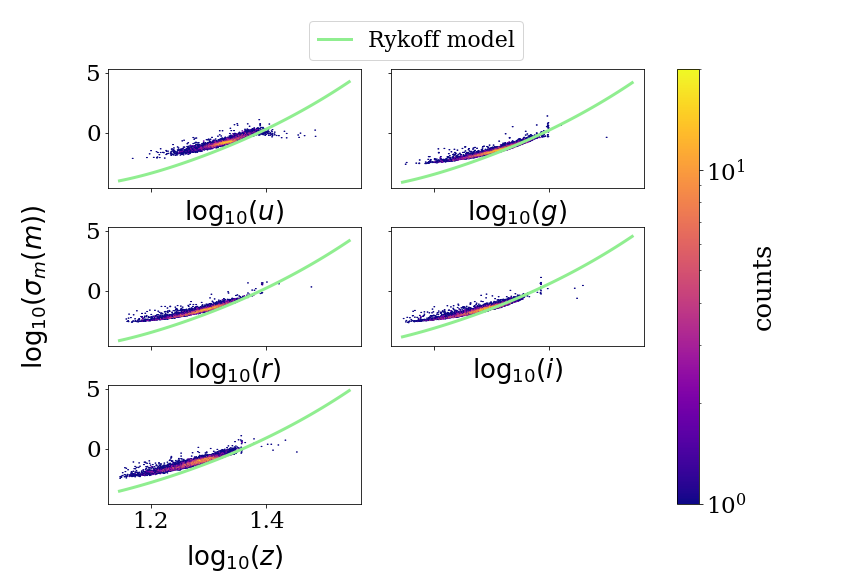}
  \caption{Magnitude uncertainties of an \textbf{SDSS-like survey}. The binned data points show the SDSS DR16 measurements \citep{Ahumada2020} of the apparent magnitudes and its corresponding uncertainties, $\sigma_m(m)$, in the $u$, $g$, $r$, $i$ and $z$ filter bands. The green lines show the respective fits to the \protect\cite{Rykoff2015} error model. We see that model and data generally match for most of the data points.}
 \label{fig:mUncertaintySDSS}
\end{figure}

\begin{figure}
 \includegraphics[width=\columnwidth]{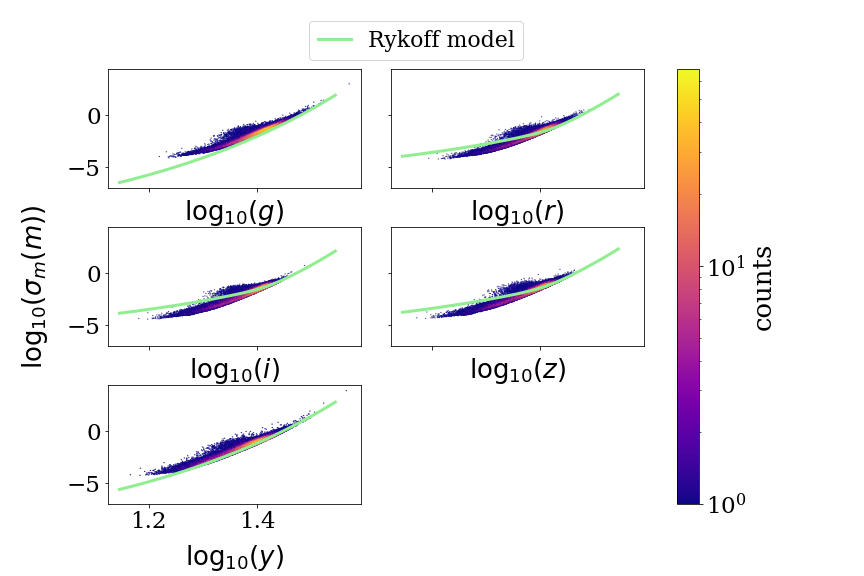}
 \caption{Magnitude uncertainties of a \textbf{HSC-like survey}. The binned data points show the HSC DR2 measurements \citep{Aihara2019} of the apparent magnitudes and its corresponding uncertainties, $\sigma_m(m)$, in the $g$, $r$, $i$, $z$ and $y$ filter bands. The green lines show the respective fits to the \protect\cite{Rykoff2015} error model. We see that model and data generally match .}
 \label{fig:mUncertaintyHSC}
\end{figure}

Figures \ref{fig:mUncertaintyDES}, \ref{fig:mUncertaintySDSS} and \ref{fig:mUncertaintyHSC} show the fitted magnitude uncertainties in each filter band and compare them to the data for every survey type, respectively.
We see that the fits agree very well with the data in the regions of high galaxy density for each survey type. In the case of DES- and HSC-like surveys, in figures \ref{fig:mUncertaintyDES} and \ref{fig:mUncertaintyHSC}, we observe that there are minor deviations of model and data in the region of small magnitudes and errors. Since the error values are very small compared to the magnitude values in this region, we do not expect these deviations to impact our simulations and analysis majorly. In the case of SDSS-like surveys in figure \ref{fig:mUncertaintySDSS}, the fits are less good compared to DES- and HSC-like surveys. However, we used the results presented in \cite{Rykoff2015} and did not perform our own fits. The largest deviations of fit and data are again in the region of small error such that they do not impact our analysis majorly.
Further, please note that our goal was not to simulate the magnitude-error correlation perfectly, but to have a model that describes the relationship sufficiently. In the plots, we can see that we achieved this goal although the given limiting magnitudes of SDSS and HSC do not correspond to $10\sigma$ magnitude limits.

Tables \ref{tab:fitDES}, \ref{tab:fitSDSS} and \ref{tab:fitHSC} summarise the fit results and the used magnitude limits.
As mentioned, we used the fit results from \cite{Rykoff2015} in the case of the SDSS-like survey.
For the DES-like survey, we performed the fit to the Y1 results and used the corresponding limiting magnitudes as described by \cite{Drlica-Wagner2018}. 
For SDSS, we compared the fit to DR16 data \citep{Ahumada2020} and used the limiting magnitudes described in \cite{Stoughton2002}.
Regarding HSC, we fitted DR2 data and used the magnitude limits of the deep survey described in \cite{Aihara2019}.

The modelled uncertainties were used to assign an error to the intrinsic magnitudes. We drew Gaussian random variables with mean zero and the standard deviations being the modelled uncertainties. We added the random values to the intrinsic magnitudes and generated the observed catalogue. 

\begin{table}
\centering
 \caption{Fit results of the magnitude uncertainty and used magnitudes limits of each filter band for a \textbf{DES-like survey}. The parameters $a$ and $b$ describe the effective exposure time in equation \ref{eq:exposureTime}. The fit was performed to DES Y1 \citep{Drlica-Wagner2018} data and the corresponding magnitudes limits were taken from \protect\cite{Drlica-Wagner2018}.}
 \label{tab:fitDES}
 \begin{tabular}{cccc}
  \hline
  Filter Band & a & b & $m_\mathrm{lim}$ \\
  \hline
  $g$ & 1.541275 & -1.000737 & 23.4 \\[3pt]
  $r$ & -0.743079 & -0.116539 & 23.2 \\[3pt]
  $i$ & 0.846085 & -1.660630 & 22.5 \\[3pt]
  $z$ & -0.044378 & -2.774705 & 21.8 \\
  \hline
 \end{tabular}
\end{table}

\begin{table}
\centering
\caption{Fit results of the magnitude uncertainty and used magnitudes limits of each filter band for an \textbf{SDSS-like survey}. The parameters $a$ and $b$ describe the effective exposure time in equation \ref{eq:exposureTime}. The fit results were taken from \protect\cite{Rykoff2015} and the corresponding magnitudes limits were taken from \protect\cite{Stoughton2002}.}
  \label{tab:fitSDSS}
  \begin{tabular}{cccc}
   \hline
   Filter Band & a & b & $m_\mathrm{lim}$ \\
   \hline
   $u$ & 3.41 & 1.15 & 22 \\[3pt]
   $g$ & 4.27 & 0.85 & 22.2 \\[3pt]
   $r$ & 4.53 & 0.91 & 22.2 \\[3pt]
   $i$ & 4.56 & 1.00 & 21.3 \\[3pt]
   $z$ & 4.39 & 1.34 & 20.5 \\
   \hline
  \end{tabular}
 \end{table}

\begin{table}
\centering
 \caption{Fit results of the magnitude uncertainty and used magnitudes limits of each filter band for an \textbf{HSC-like survey}. The parameters $a$ and $b$ describe the effective exposure time in equation \ref{eq:exposureTime}. The fit was performed to HSC DR2 \citep{Aihara2019} data and the corresponding magnitudes limits were taken from \protect\cite{Aihara2019}.}
 \label{tab:fitHSC}
 \begin{tabular}{cccc}
  \hline
  Filter Band & a & b & $m_\mathrm{lim}$ \\
  \hline
  $g$ & 135.095407 & 20.425863 & 27.8 \\[3pt]
  $r$ & -2.270331 & 0.924841 & 27.4 \\[3pt]
  $i$ & -0.272460 & 0.563884 & 27.1 \\[3pt]
  $z$ & -0.707218 & 0.618222 & 26.6 \\[3pt]
  $y$ & 61.117607 & 13.634286 & 25.6 \\
  \hline
 \end{tabular}
\end{table}

\subsubsection{Magnitude Cuts}
\label{sec:magnitude_cuts}
To generate mock observed catalogues, we needed to account for further observational effects.
Besides environmental factors like weather, which we did not include, signal-to-noise effects had to be addressed in our simulations. 
Taking observations, only measurements that exceed a certain signal-to-noise threshold are kept, which guarantees the quality of the data. Fainter sources are more affected than brighter objects because the noise does not linearly scale with the brightness.
The signal-to-noise ratio defines how deep a survey is and sets a lower limit on the brightness of an object to be observed.

However, signal-to-noise selection is a complex process and mocking it is not straightforward. Simulating realistic signal-to-noise for galaxies needs to account for large sets of galaxy properties (such as luminosity, size, profile and ellipticity) and how these correlate with the point-spread-function properties (size and shape) at the positions of galaxies \citep{Berge2013, Herbel2017}. This means that we would need to implement realistic image simulations to fully account for signal-to-noise selection effects, which exceeds the purpose of our paper. Therefore, we have focused on a fixed magnitude cut where the sample should be complete.

In our work, we chose a cut in the $i$-band. That means all objects whose $i$-band magnitude is fainter than a certain cut-off point $i_\mathrm{upper}$ were removed.
As the investigated surveys have varying depths, the respective upper-cut magnitudes were different. In the case of a DES-like survey, we chose $i_\mathrm{upper, DES} = 24$ and for an HSC-like survey, we used $i_\mathrm{upper, HSC} = 27$.
In the case of SDSS, $i_\mathrm{upper, SDSS}$ were 21.

The saturation of pixels is another crucial effect that occurs during observations.
If a source is too bright, too many photons are detected by the CCD and pixels start to saturate. To account for this, we introduced a magnitude cut for bright objects. 
Again, we used a cut in the $i$-band. However, the cut was the same for each survey. We removed all objects with $i$-band magnitudes brighter than  $i_\mathrm{lower} = 18$.
Table \ref{tab:cuts} summarises all cuts that were used in our work.

\begin{table}
\centering
 \caption{The upper and lower magnitude limits that were used in our analysis. Only objects with $i$-band magnitudes between $i_\mathrm{lower}$ and $i_\mathrm{upper}$ were kept in the catalogues.}
 \label{tab:cuts}
 \begin{tabular}{ccc}
  \hline
  Survey & $i_\mathrm{upper}$ & $i_\mathrm{lower}$\\
  \hline
  DES & 24 & 18  \\[3pt]
  SDSS & 21 & 18  \\[3pt]
  HSC & 27 & 18  \\
  \hline
 \end{tabular}
\end{table}

\section{Method}
\label{sec:method}
In this section, we describe the method that we used to investigate the sensitivity of the redshift distribution to the Schechter parameters and the impact of those parameters to the observables.

First, we simulated the intrinsic fiducial galaxy catalogues for all surveys. To do this, we used the SkyPy Python package, which enables us to perform the simulations according to the model described in section \ref{sec:model}.
The intrinsic catalogues contained redshift values, absolute Johnson B-band magnitudes and apparent magnitudes in different filter bands specific to the survey.
We summarised the fiducial values of the parameters in sections \ref{sec:schechter_params} and \ref{sec:dirichlet_params}.

Afterwards, the intrinsic catalogues were transformed into observed catalogues. We added an observational error to the apparent magnitudes and performed magnitude cuts to remove too faint and too bright objects.
We refer to section \ref{sec:surveys} for further details.

To analyse the sensitivity of the surveys to the different Schechter parameters, we changed the value of one of the parameters by adding and subtracting its error and repeated the whole simulation process as described in the previous paragraphs. The positive and negative errors correspond to the 84th and 50th percentile values, respectively.
Note, the sensitivity is relative to the ability of modern datasets to constrain the parameters (see table \ref{tab:dirichletParams}).
Modifying the parameter changed the luminosity function and the simulation resulted
in a galaxy catalogue with different redshift, magnitude and colour distributions. We call this catalogue the ``test'' catalogue.
Remember that the sample is defined by the Schechter parameters and the Dirichlet coefficients. We concentrated on the effect of the Schechter parameters on the redshift distribution. The correlation between the parameters and coefficients, if there is one, has not been measured in the literature and determining this correlation is beyond the scope of our paper.

Comparing the redshift distributions of fiducial and test catalogue allowed us to identify the sensitivity of the survey's redshift distribution to the modified parameter. By comparing the colour and magnitude distributions of fiducial and test catalogue, we diagnosed different observables (magnitude or colour distribution) that are sensitive to a change in the parameter value and, thus, could be used to constrain the parameters.

In this context, ``sensitive to a Schechter parameter'' means that the redshift distribution of the galaxy catalogue changes after modifying the parameter. 
Furthermore, a greater change indicates a higher sensitivity of this survey to the parameter.
Accordingly, an observable has better constraining ability if it is more affected by the parameter change.

\subsection{Investigating Redshift Sensitivity}
\label{sec:redshift_investigation}
Since the mean of the redshift distribution is the leading order term in weak lensing analyses \citep{Tessore2020}, we concentrated on the change of the mean redshift.
Comparing the mean redshift of the fiducial and the test redshift distributions is beneficial in several ways.
First, one can investigate the absolute change that informs about the required precision of the parameter.
Second, the relative change reveals the parameter that has the strongest impact on the redshift distribution.

We define the relative change as
\begin{equation}
\label{eq:mean_redshift_comparison}
    \delta z_{\mathrm{mean}} = \frac{|\Delta z_{\mathrm{mean}}|}{\overline{z}_{\mathrm{fid}}} \;,
\end{equation}
where $\Delta z_{\mathrm{mean}} = \overline{z}_{\mathrm{test}} - \overline{z}_{\mathrm{fid}}$ is the absolute change of the mean redshift, $\overline{z}_{\mathrm{test}}$ is the mean redshift of the test catalogue and $\overline{z}_{\mathrm{fid}}$ is the mean redshift of the fiducial catalogue.

As the simulations are affected by statistical fluctuations, we made sure that the change of the mean redshift did not arise from statistical effects.
To account for this, we generated 100 realisations of the fiducial and test catalogues.
We then compared the mean redshifts by looking at the mean of $\delta z_{\mathrm{mean}}$ and $\Delta z_{\mathrm{mean}}$ over the 100 simulations. Hence, $\overline{\delta z}_{\mathrm{mean}}$ and $\overline{\Delta z}_{\mathrm{mean}}$ were more appropriate to quantify the change in the redshift distribution.

\subsection{Important Observables}
After identifying the parameters with the strongest impact on the redshift distribution, we focused on how the Schechter parameters affect the observables. Thus, we could find the observables that are most sensitive to the parameters and that could help constraining these parameters.

In our analysis, we compared the magnitude distributions and colour distributions of fiducial and test catalogues by using the Anderson-Darling (AD) test \citep{Scholz1987, Feigelson2012}. 
For the test, we used the assumption that the two tested samples are drawn from the same distribution as the null hypothesis.
The AD test is a statistical test measuring the sum of the squared differences of the distributions, $d_\mathrm{AD}$. The sensitivity of the corresponding observable to the parameter increases with growing value of $d_\mathrm{AD}$.
The relating p-value describes the probability to obtain a test statistic $d_\mathrm{AD}$ at least as extreme as measured under the assumption that the null hypothesis is true.

As described in section \ref{sec:redshift_investigation}, we ran 100 simulations to account for numerical noise. We calculated the 95-th percentiles of the 100 d-values, $d_{\mathrm{AD},95}$, and p-values, $p_{\mathrm{AD},95}$, for each observable. 
These values then indicated the change of the observable after changing the Schechter parameter. Furthermore, we looked for large values of $d_{\mathrm{AD},95}$ and small values of $p_{\mathrm{AD},95}$.

We repeated the whole process of simulating the test catalogues and comparing them to the fiducial catalogues for all eight Schechter parameters and all surveys. 
We performed the analysis after changing each parameter value by its given positive and negative error.
That enabled us to identify which survey is sensitive to which parameters. In addition to this, we could determine which observables and which surveys show higher sensitivity to the specific parameters and can, therefore, be more useful in constraining the parameters.

\section{Results}
\label{sec:results}
This section describes the results of our analysis.
We start with identifying the parameters that have the strongest effect on the redshift distribution of each survey in section \ref{sec:redshift_sensitivity}.
We finalise with the discussion about the impact of the Schechter parameters on the different observables in section \ref{sec:important_observables}.

\subsection{Redshift Sensitivity}
\label{sec:redshift_sensitivity}
As described in section \ref{sec:method}, we compared the redshift distributions of fiducial and test catalogue using sets of 100 simulations.
The mean of the relative change, $\overline{\delta z}_{\mathrm{mean}}$, and the mean of the absolute change, $\overline{\Delta z}_{\mathrm{mean}}$, of the mean redshift indicated the parameters with the highest impact on each survey.
Tables \ref{tab:result_des}, \ref{tab:result_hsc} and \ref{tab:result_sdss} summarise our results of the relative and absolute mean redshift change. 
They show the impact on the redshift distribution after updating each parameter in positive and negative directions.

\begin{table*}
\centering
\caption{Relative change, $\overline{\delta z}_{\mathrm{mean}}$, and absolute change, $\overline{\Delta z}_{\mathrm{mean}}$, of the mean redshift for a \textbf{DES-like survey}. The values correspond to the mean of 100 realisations. They indicate the change in the redshift distribution after increasing (Positive) and decreasing (Negative) the corresponding parameter by its given errors. Higher values of $\overline{\delta z}_{\mathrm{mean}}$ indicate stronger sensitivity to the corresponding parameter. $\overline{\Delta z}_{\mathrm{mean}}$ is a measure for the required precision of the mean redshift.}
\label{tab:result_des}
\begin{tabular}{ccccc}
\hline
    \multirow{2}{*}{Parameter} & \multicolumn{2}{c}{Positive}    & \multicolumn{2}{c}{Negative}   \\ \cline{2-5} \\[-1.5ex]
    & \multicolumn{1}{c}{$\overline{\delta z}_{\mathrm{mean}}$ in \%} & $\overline{\Delta z}_{\mathrm{mean}}$ & \multicolumn{1}{c}{$\overline{\delta z}_{\mathrm{mean}}$ in \%} & $\overline{\Delta z}_{\mathrm{mean}}$ \\ \hline
    $a_\mathrm{M, SF}$ & \multicolumn{1}{c}{10.5} & -0.071 & \multicolumn{1}{c}{44.9} & 0.304 \\ [3pt]
    $a_\mathrm{M, Q}$ & \multicolumn{1}{c}{3.8} & -0.026 & \multicolumn{1}{c}{13.5} & 0.091 \\ [3pt]
    $a_{\phi, \mathrm{SF}}$ & \multicolumn{1}{c}{4.9} & 0.033 & \multicolumn{1}{c}{3.6} &  -0.024\\ [3pt]
    $a_{\phi, \mathrm{Q}}$ & \multicolumn{1}{c}{8.7} & 0.059 & \multicolumn{1}{c}{2.8} & -0.019 \\ [3pt]
    $b_\mathrm{M, SF}$ & \multicolumn{1}{c}{5.5} & -0.037 & \multicolumn{1}{c}{7.7} & 0.053 \\ [3pt]
    $b_\mathrm{M, Q}$ & \multicolumn{1}{c}{2.7} & -0.018 & \multicolumn{1}{c}{3.5} & 0.024 \\ [3pt]
    $b_{\phi, \mathrm{SF}}$ & \multicolumn{1}{c}{0.5} & -0.004 & \multicolumn{1}{c}{1.0} & 0.007 \\[3pt] 
    $b_{\phi, \mathrm{Q}}$ & \multicolumn{1}{c}{1.1} & 0.008 & \multicolumn{1}{c}{1.7} & -0.012 \\ \hline
\end{tabular}
\end{table*}

Table \ref{tab:result_des} summarises the sensitivity of DES-like surveys.
We found that a DES-like survey is most sensitive to $a_{\mathrm{M, SF}}$. Decreasing $a_{\mathrm{M, SF}}$ increases the mean redshift by ${\sim} 45$ per cent. This relative change corresponds to an absolute change of 0.304. 
The rest of the parameters have less impact on the mean redshift. However, $a_{\mathrm{M, Q}}$, $a_{\phi,\mathrm{Q}}$ and $b_{\mathrm{M, SF}}$ can also have a significant effect on the redshift distribution.
Updating these parameters causes relative changes of the mean redshift up to 13.5 per cent, which corresponds to an absolute change of 0.091. 

\begin{table*}
\centering
 \caption{Relative change, $\overline{\delta z}_{\mathrm{mean}}$, and absolute change, $\overline{\Delta z}_{\mathrm{mean}}$, of the mean redshift for an \textbf{HSC-like survey}. The values correspond to the mean of 100 realisations. They indicate the change in the redshift distribution after increasing (Positive) and decreasing (Negative) the corresponding parameter by its given errors. Higher values of $\overline{\delta z}_{\mathrm{mean}}$ indicate stronger sensitivity to the corresponding parameter. $\overline{\Delta z}_{\mathrm{mean}}$ is a measure for the required precision of the mean redshift.}
 \label{tab:result_hsc}
\begin{tabular}{ccccc}
\hline
    \multirow{2}{*}{Parameter} & \multicolumn{2}{c}{Positive}    & \multicolumn{2}{c}{Negative}   \\ \cline{2-5} \\[-1.5ex]
    & \multicolumn{1}{c}{$\overline{\delta z}_{\mathrm{mean}}$ in \%} & $\overline{\Delta z}_{\mathrm{mean}}$ & \multicolumn{1}{c}{$\overline{\delta z}_{\mathrm{mean}}$ in \%} & $\overline{\Delta z}_{\mathrm{mean}}$ \\ \hline
    $a_\mathrm{M, SF}$ & \multicolumn{1}{c}{17.63} & -0.2053 & \multicolumn{1}{c}{22.69} & 0.2643 \\ [3pt]
    $a_\mathrm{M, Q}$ & \multicolumn{1}{c}{1.50} & -0.0175 & \multicolumn{1}{c}{1.74} & 0.0203 \\ [3pt]
    $a_{\phi, \mathrm{SF}}$ & \multicolumn{1}{c}{12.37} & 0.1441 & \multicolumn{1}{c}{9.29} &  -0.1082\\ [3pt]
    $a_{\phi, \mathrm{Q}}$ & \multicolumn{1}{c}{8.68} & 0.1011 & \multicolumn{1}{c}{1.18} & -0.0138 \\ [3pt]
    $b_\mathrm{M, SF}$ & \multicolumn{1}{c}{4.19} & -0.0488 & \multicolumn{1}{c}{3.84} & 0.0447 \\ [3pt]
    $b_\mathrm{M, Q}$ & \multicolumn{1}{c}{0.55} & -0.0064 & \multicolumn{1}{c}{0.51} & 0.0059 \\ [3pt]
    $b_{\phi, \mathrm{SF}}$ & \multicolumn{1}{c}{0.05} & -0.0006 & \multicolumn{1}{c}{0.06} & 0.0007 \\[3pt] 
    $b_{\phi, \mathrm{Q}}$ & \multicolumn{1}{c}{0.06} & 0.0007 & \multicolumn{1}{c}{0.13} & -0.0015 \\ \hline
\end{tabular}
\end{table*}

Table \ref{tab:result_hsc} shows the results for an HSC-like survey.
Again, $a_\mathrm{M, SF}$ has the strongest impact on the redshift distribution. The mean redshift of an HSC-like survey changes by ${\sim}23$ per cent after decreasing $a_\mathrm{M, SF}$ by its negative error. Therefore, the uncertainty of $a_\mathrm{M, SF}$ shifts the simulated mean redshift by up to ${\sim} 0.26$.
Furthermore, we found that the parameters $a_{\phi, \mathrm{SF}}$ and $a_{\phi, \mathrm{Q}}$ also have strong impact on the mean redshift. Increasing these parameters causes relative changes of ${\sim} 12.4$ per cent and ${\sim} 8.7$ per cent, which corresponds to absolute changes of about 0.14 and 0.10, respectively.

\begin{table*}
\centering
 \caption{Relative change, $\overline{\delta z}_{\mathrm{mean}}$, and absolute change, $\overline{\Delta z}_{\mathrm{mean}}$, of the mean redshift for an \textbf{SDSS-like survey}. The values correspond to the mean of 100 realisations. They indicate the change in the redshift distribution after increasing (Positive) and decreasing (Negative) the corresponding parameter by its given errors. Higher values of $\overline{\delta z}_{\mathrm{mean}}$ indicate stronger sensitivity to the corresponding parameter. $\overline{\Delta z}_{\mathrm{mean}}$ is a measure for the required precision of the mean redshift.}
 \label{tab:result_sdss}
\begin{tabular}{ccccc}
\hline
    \multirow{2}{*}{Parameter} & \multicolumn{2}{c}{Positive}    & \multicolumn{2}{c}{Negative}   \\ \cline{2-5} \\[-1.5ex]
    & \multicolumn{1}{c}{$\overline{\delta z}_{\mathrm{mean}}$ in \%} & $\overline{\Delta z}_{\mathrm{mean}}$ & \multicolumn{1}{c}{$\overline{\delta z}_{\mathrm{mean}}$ in \%} & $\overline{\Delta z}_{\mathrm{mean}}$ \\ \hline
    $a_\mathrm{M, SF}$ & \multicolumn{1}{c}{3.8} & -0.013 & \multicolumn{1}{c}{9.1} & 0.031 \\ [3pt]
    $a_\mathrm{M, Q}$ & \multicolumn{1}{c}{5.8} & -0.020 & \multicolumn{1}{c}{10.5} & 0.036 \\ [3pt]
    $a_{\phi, \mathrm{SF}}$ & \multicolumn{1}{c}{0.9} & 0.003 & \multicolumn{1}{c}{0.7} &  -0.002\\ [3pt]
    $a_{\phi, \mathrm{Q}}$ & \multicolumn{1}{c}{4.8} & 0.016 & \multicolumn{1}{c}{3.3} & -0.011 \\ [3pt]
    $b_\mathrm{M, SF}$ & \multicolumn{1}{c}{2.7} & -0.009 & \multicolumn{1}{c}{5.9} & 0.020 \\ [3pt]
    $b_\mathrm{M, Q}$ & \multicolumn{1}{c}{7.1} & -0.024 & \multicolumn{1}{c}{9.0} & 0.031 \\ [3pt]
    $b_{\phi, \mathrm{SF}}$ & \multicolumn{1}{c}{1.5} & -0.005 & \multicolumn{1}{c}{2.4} & 0.008 \\[3pt] 
    $b_{\phi, \mathrm{Q}}$ & \multicolumn{1}{c}{2.7} & 0.009 & \multicolumn{1}{c}{5.4} & -0.019 \\ \hline
\end{tabular}
\end{table*}

Finally, we focused on the redshift sensitivity of SDSS-like surveys.
Table \ref{tab:result_sdss} presents the impact of the Schechter parameters on an SDSS-like redshift distribution.
The parameter $a_\mathrm{M, Q}$ has the strongest effect on the redshift distribution. The mean redshift changes by 10.5 per cent after decreasing $a_\mathrm{M, Q}$. The corresponding absolute change of 0.036 illustrates that SDSS-like surveys are not strongly affected by the parameter uncertainties. 

Figures \ref{fig:all_redshift_des}, \ref{fig:all_redshift_hsc} and \ref{fig:all_redshift_sdss} in appendix \ref{sec:appendix_redshift} illustrate the differences in the fiducial and test redshift distributions for all survey types.
The plots show the redshift distributions of which the summary statistics are given in tables \ref{tab:result_des}, \ref{tab:result_hsc} and \ref{tab:result_sdss}. Larger values of $\overline{\delta z}_{\mathrm{mean}}$ correspond to visually larger differences in the distributions. 

\subsubsection{Survey Comparison}
\label{sec:redshift_survey_comparison}

\begin{figure*}
    \includegraphics[width=2\columnwidth]{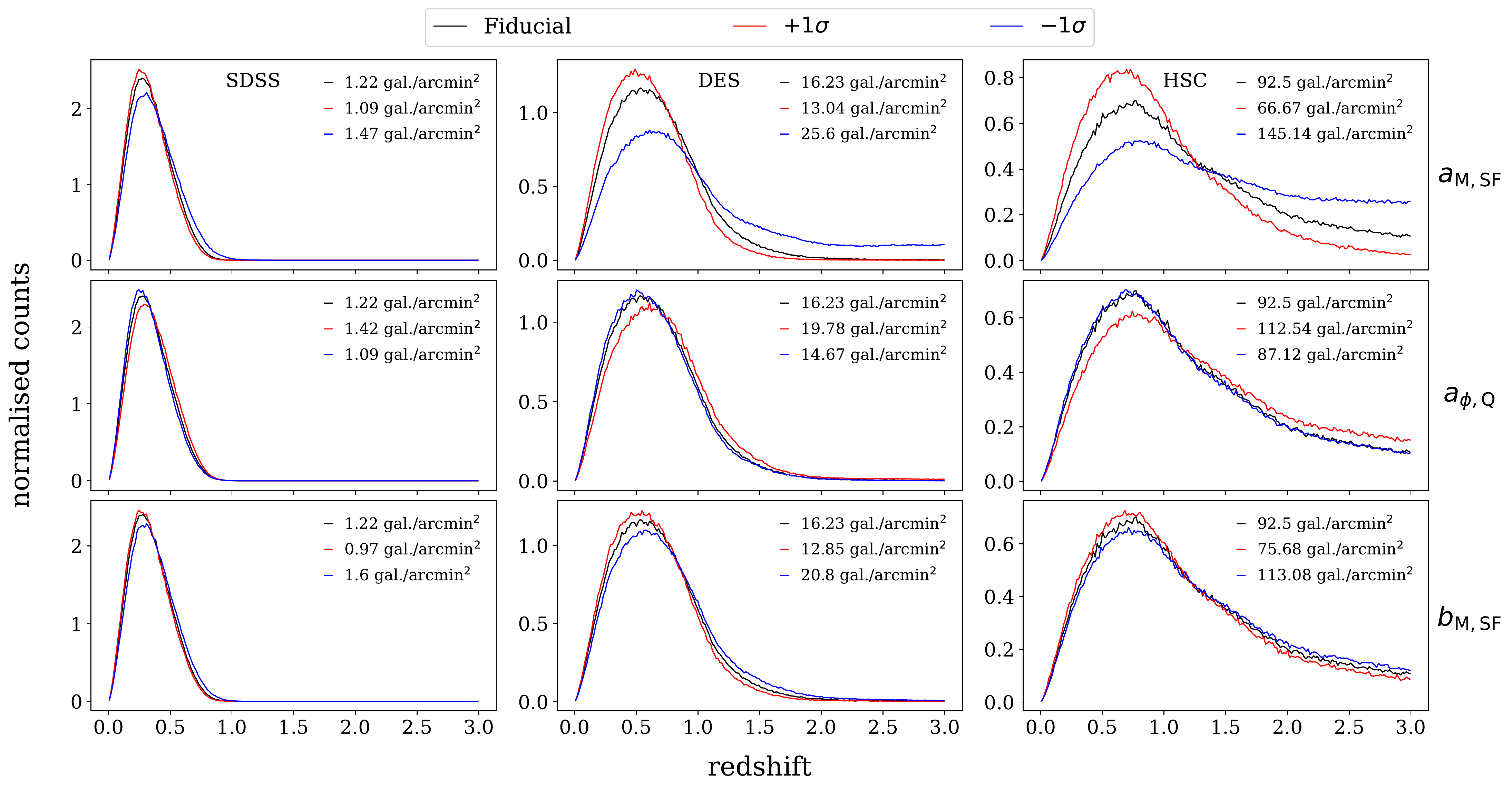}
    \caption{Comparison of the normalised simulated fiducial (black) and test redshift distributions after increasing (red) and decreasing (blue) $a_\mathrm{M, SF}$, $a_{\phi, \mathrm{Q}}$ and $b_\mathrm{M, SF}$ by $1\sigma$. Note that the distributions are normalised such that the integral over the redshift range $z \in [0,3]$ is one. The columns correspond to the different survey types. The first column shows the results of SDSS-like surveys. The second and third columns display the results of DES- and HSC-like surveys, respectively. The rows correspond to the different parameters. Furthermore, we indicate the number of galaxies per square arcminute in the legends of each plot. We see that DES- and HSC-like surveys are more sensitive to $a_\mathrm{M, SF}$, $a_{\phi, \mathrm{Q}}$ and $b_\mathrm{M, SF}$. Furthermore, $a_\mathrm{M, SF}$ has the strongest impact.}
    \label{fig:redshift_comparison}
\end{figure*}

Figure \ref{fig:redshift_comparison} compares the fiducial and test redshift distributions for all survey types after modifying different parameters. Each plot shows the fiducial redshift distribution in black, the test redshift distribution after decreasing the corresponding parameter by $1\sigma$ in blue and the test redshift distribution after increasing the parameter by $1\sigma$ in red.
The columns correspond to the different surveys, the first column shows the results for an SDSS-like survey, the second for a DES-like and the third for an HSC-like survey.
Moreover, we compare the effect of three different parameters in the individual rows. The first row shows the impact of $a_\mathrm{M, SF}$, the second of $a_{\phi, \mathrm{Q}}$ and the third of $b_\mathrm{M, SF}$.
In addition, we indicate the number of galaxies per square arcminute in the legend of each plot.

The first row highlights that $a_\mathrm{M, SF}$ has a higher impact on DES- and HSC-like surveys than on SDSS-like surveys. Decreasing $a_\mathrm{M, SF}$ increases the mean redshift. Increasing $a_\mathrm{M, SF}$ decreases the mean redshift. This effect also correlates with the galaxy number density.
The second row shows similar characteristics. DES- and HSC-like surveys are more sensitive to $a_{\phi, \mathrm{Q}}$ and increasing the parameter has a stronger effect than decreasing it.
Especially, the plots indicate that a change of the parameters that increases the mean redshift has generally a stronger effect than a change that decreases the mean redshift. 
This attribute is independent of the direction of the parameter change and follows from the fact that fewer high-redshift than low-redshift galaxies are measured. The change that increases the mean redshift is also always accompanied by an increase in the number density. Hence, constraining the direction of change that is causing the increase in the mean redshift is more important than the direction that decreases the mean redshift.

Looking at the model of the characteristic magnitude (eq. \ref{eq:Mstar}), the above-mentioned observations agree with theoretical expectations. As $a_\mathrm{M}$ describes the redshift evolution of the characteristic magnitude, variations of $a_\mathrm{M}$ directly change the bright end of the luminosity function. Since galaxy observations are in general limited by the flux, the changes in the bright end influence the abundance of measured high redshift objects.
Although $b_\mathrm{M}$ is also affecting the characteristic magnitude it does not affect the redshift evolution directly and is expected to have less impact on the redshift distribution than $a_\mathrm{M}$.

The different survey types also have distinct sensitivity to the parameters of the two galaxy populations.
Tables \ref{tab:result_des}, \ref{tab:result_hsc} and \ref{tab:result_sdss} show that DES- and HSC-like surveys tend to be more sensitive to parameters of the star-forming population.
On the other hand, SDSS-like surveys are more sensitive to the parameters of quenched galaxies.
For DES- and HSC-like surveys, $a_\mathrm{M, SF}$ has the strongest impact on the redshift distribution. Conversely, the mean redshift of an SDSS-like survey is most sensitive to $a_\mathrm{M, Q}$ and $b_\mathrm{M, Q}$. 
The difference in the sensitivities follows from the variety in the depths of the surveys. An SDSS-like survey has a lower magnitude limit than the other surveys and measures mostly closer galaxies, which have lower redshift values. These galaxies had more time to get quenched and the ratio of star-forming to quenched galaxies is smaller than for DES- and HSC-like surveys.
DES- and HSC-like surveys are more sensitive to parameters of the star-forming galaxies because these survey types also measure younger galaxies. However, the majority of galaxies measured by these surveys still have low redshifts. 
That explains why parameters of the quenched galaxies can have a strong impact on the redshift distribution of DES- and HSC-like surveys, for example, the 13.5 per cent change of the DES-like mean redshift or the 17.5 per cent change in the case of an HSC-like survey after decreasing $a_\mathrm{M, Q}$.

To summarise the analysis of the sensitivity to the Schechter parameters, we found that $a_\mathrm{M, SF}$ has the highest impact on the redshift distribution, although the sensitivity depends on the survey type. We conclude that $a_\mathrm{M, SF}$ is the most important parameter to constrain. Within the given uncertainty of $a_\mathrm{M, SF}$ the modelled mean redshift can change by up to 0.3. This can cause strong bias in the cosmological constraints when using weak lensing observations and the forward modelled redshift distribution.
We present the most sensitive observables in the following section.

\subsection{Colour and Magnitude Distributions}
\label{sec:important_observables}

\begin{figure*}
    \includegraphics[width=1.5\columnwidth]{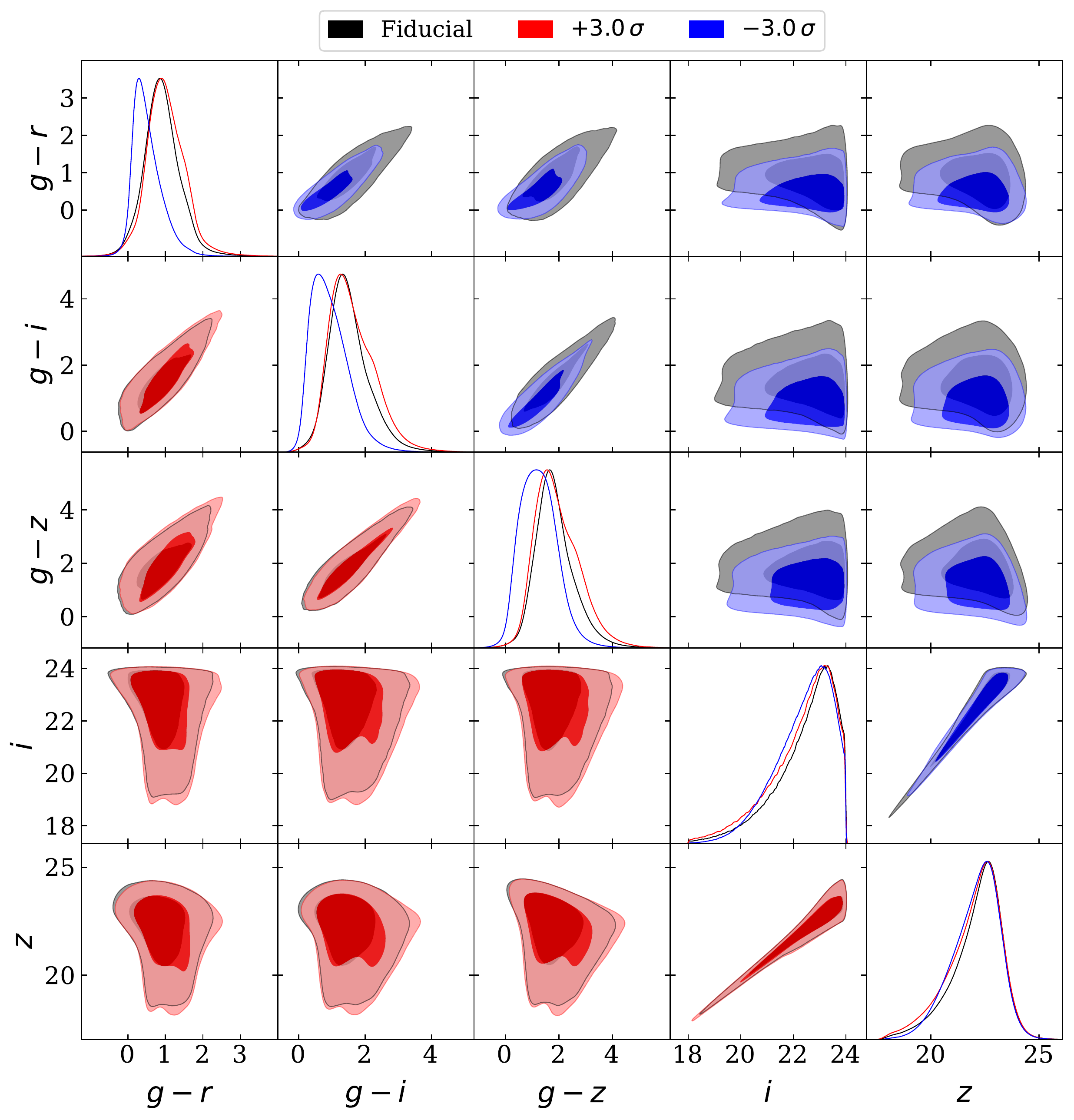}
    \caption{Corner plot of the $g - r$, $g - i$, $g - z$ colour and $i$- and $z$-band magnitude distributions for a \textbf{DES-like survey}. Black lines and contours correspond to the distributions in the fiducial catalogue. The red and blue colours represent the test catalogues after increasing and decreasing $a_\mathrm{M, SF}$ by $3\sigma$, respectively. The colour distributions are more affected by a change of $a_\mathrm{M, SF}$. A combination of colour and magnitude is most suitable to constrain $a_\mathrm{M, SF}$.}
    \label{fig:observable_comparison}
\end{figure*}

To investigate the effect of the parameter changes on the observables, we compared the magnitude and colour distributions of the fiducial and test catalogues.
We focused on DES-like surveys because their depths are similar to those of future and upcoming surveys as previously mentioned. We investigated the effect on all observables (apparent magnitudes in the different bands and the corresponding colours) but concentrated on the five most sensitive observables to $a_\mathrm{M, SF}$, which are the $g - r$, $g - i$, $g - z$ colour and $i$- and $z$-band magnitude distributions.

We focused on the results for $a_\mathrm{M, SF}$ for two reasons. First, we showed that $a_\mathrm{M, SF}$ causes the strongest effect and is most important to constrain.
Second, we found that the to $a_\mathrm{M, SF}$ most sensitive observables are generally sensitive to the other parameters as well. Especially, the colour distributions are very sensitive to changes in most of the parameters\footnote{That is even true for the other survey types.}.

We found that the colour distributions $g - r$, $g - i$, $g - z$ and the magnitude distributions in the $i$- and $z$-band are the five most sensitive observables to $a_\mathrm{M, SF}$ and have the highest potential to help constraining its value in the future.
As described earlier, we compared the colour and magnitude distributions of fiducial and test catalogues quantitatively using the Anderson-Darling test. In the case of the colour distributions $g - r$, $g - i$ and $g - z$, we obtained d-values of the order $10^4$ after decreasing the parameter value. In the case of the magnitude distributions in the $i$- and $z$-bands, the d-values were of the order $10^3$. Hence, the colour distributions are more sensitive to a modification of $a_\mathrm{M, SF}$ than the magnitude distributions. The d-value after decreasing the value of $a_\mathrm{M, SF}$ by its error for the $g - r$ colour distributions was ${\sim}21,300$. This value was the largest in our analysis.

Figure \ref{fig:observable_comparison} shows a corner plot comparing these five observables. The plot contains the one-dimensional distributions and contours of all observables. The black lines and contours correspond to the fiducial catalogue. The colours blue and red represent the results of the test catalogue after decreasing and increasing $a_\mathrm{M, SF}$ by $3\sigma$.
Again, blue corresponds to a negative and red to a positive change.

The one-dimensional distributions in figure \ref{fig:observable_comparison} verify the results indicated by the AD test that the colours are more sensitive to a change of $a_\mathrm{M, SF}$ than the magnitudes.
Once more, we observed that the negative change, which increases the mean redshift, has a stronger impact on the distributions than a positive change.
The figure also shows that modifying the parameter affects the shape and the peak of the colour distributions. However, in the case of the magnitude distributions, the shapes are affected by the parameter change but the peaks stay the same. For different colour and magnitude choices in this comparison, the conclusion might be different because of the distinct sensitivity in the single bands. 

The contours show that a combination of colour and magnitude is generally most sensitive to the parameters and might be the best choice to constrain them. The differences in the fiducial and test contours appear larger in colour-magnitude space.
In addition, the results indicate that the magnitude-magnitude distributions are less sensitive to the parameters than the distributions in colour-colour space. However, the sensitivity of the colour-colour distributions is still smaller than in colour-magnitude space.
Additionally, the choice of colour-magnitude combination does not influence the sensitivity to the parameter change because all test contours feature similar deviations from the fiducial contours.
Note that our analysis showed that these colour-magnitude pairs are sensitive to the rest of the parameters as well and not only $a_\mathrm{M, SF}$. The AD tests indicated high sensitivity of these observables to all parameters. In general, the contour plots for the other parameters are similar to figure \ref{fig:observable_comparison} and only small deviations exist. Therefore, these pairs might also be able to help constraining the other parameters.
Furthermore, our analysis has shown the same results if we chose colour-magnitude pairs that are not included in the figure (e.g $i-z$ or $r-i$). But, as mentioned, we included the colours and magnitudes with the highest sensitivity.

\begin{figure*}
    \includegraphics[width=1.5\columnwidth]{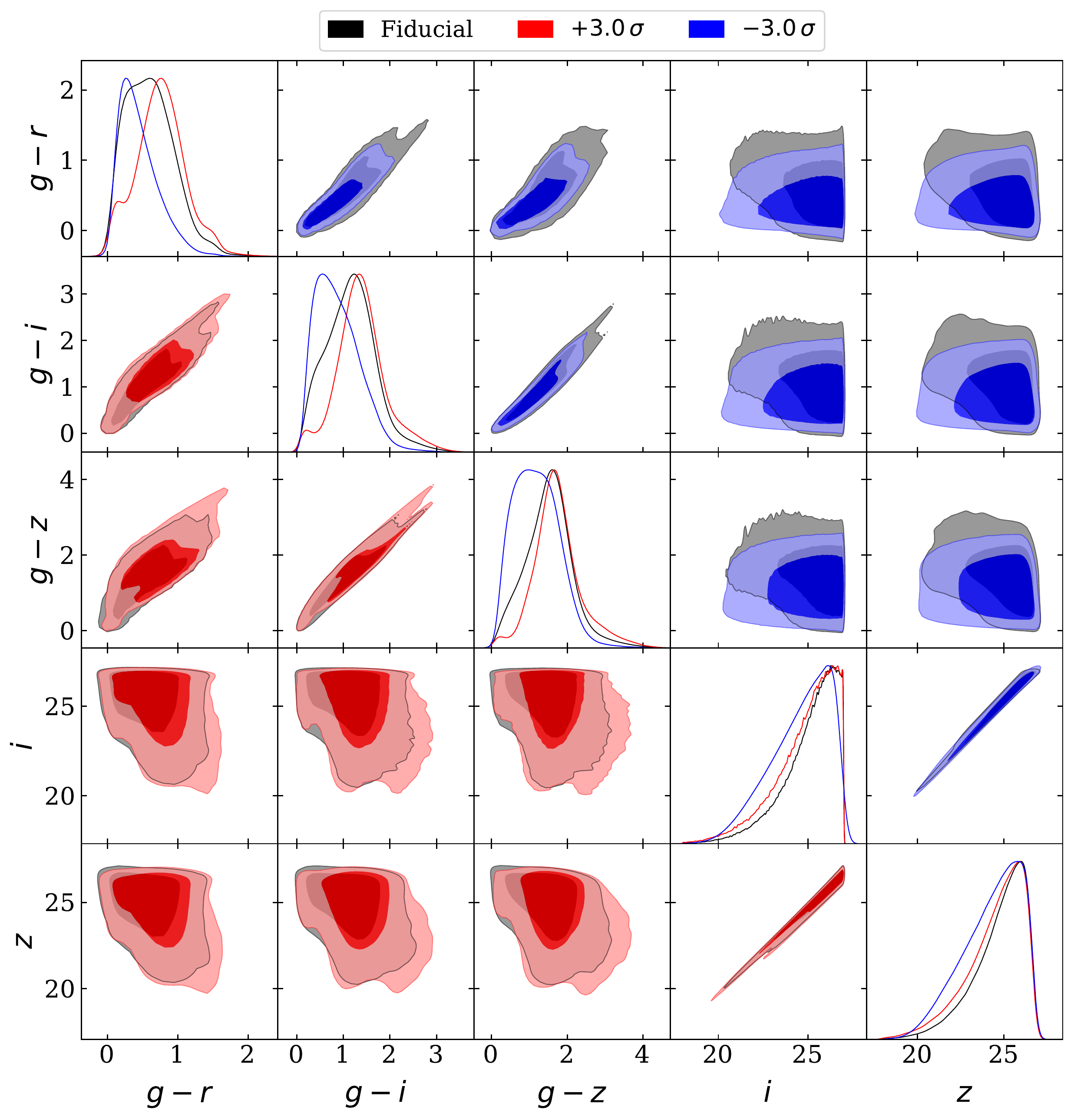}
    \caption{Corner plot of the $g - r$, $g - i$, $g - z$ colour and $i$- and $z$-band magnitude distributions for an \textbf{HSC-like survey}. Black lines and contours correspond to the distributions in the fiducial catalogue. The red and blue colours represent the test catalogues after increasing and decreasing $a_\mathrm{M, SF}$ by $3\sigma$, respectively. Comparing this plot with figure \ref{fig:observable_comparison}, we see that an HSC-like survey is strongly affected by the parameter change than a DES-like survey.}
    \label{fig:observable_comparison_hsc}
\end{figure*}

We also investigated the effect of modifying the Schechter parameters to HSC-like measurements. Figure \ref{fig:observable_comparison_hsc} shows the same plot as figure \ref{fig:observable_comparison} but for an HSC-like survey.
Comparing the two plots illustrates that a deep HSC-like survey has higher sensitivity and therefore better constraining ability than a DES-like survey. The HSC-like contours of fiducial and test catalogues have a larger deviation than in the case of DES-like surveys.
Furthermore, the one-dimensional colour and magnitude distributions are more responsive to the parameter modification. Although the d-values for a negative modification of the parameter are similar to the results of a DES-like survey, the d-values after adding the parameter error to its value are larger in the case of HSC-like measurements.
The d-values of the colour distributions $g - r$, $g - i$ and $g - z$ are ${\sim} 6\cdot 10^3$ in the case of HSC-like survey and, hence, about six times larger than in the case of DES-like surveys.

In summary, we suggest that a combination of colour-magnitude measurements is more sensitive to changes to the Schechter parameters and can be more useful to constrain them. Thereby, the exact selection of colour-magnitude pair has a minor role because the sensitivity is in general similar for all combinations.
Additionally, HSC-like measurements can improve the constraining power compared to DES-like measurements.

\section{Conclusions}
\label{sec:conclusion}
Different cosmological probes are used to constrain the parameters of dark energy as well as dark matter and, therefore, to improve the cosmological model. 
One of these probes is weak gravitational lensing. Cosmic shear measurements rely on knowing the redshift distribution $n(z)$ of the galaxies under study. However, measuring $n(z)$ with sufficient precision is difficult. Spectroscopic redshifts are most accurate but current and upcoming experiments are too large to measure redshifts efficiently using spectroscopy.
On the other hand, photometric redshifts do not have adequate precision.

\cite{Herbel2017} used a forward modelling approach to get a full posterior of $n(z)$ without having information about each individual redshift. Their model is based on redshift-dependent Schechter luminosity functions (see eq. \ref{eq:schechter}).
In this model, the characteristic magnitude is a linear function of redshift $z$ and the amplitude has an exponential relation to $z$.
The model includes two different galaxy types, star-forming (SF) galaxies and quenched (Q) galaxies.
Therefore, the number of galaxies per comoving volume and absolute magnitude depends on the eight Schechter luminosity function parameters $a_{\mathrm{M, SF}}$, $a_{\mathrm{M, Q}}$, $a_{\phi, \mathrm{SF}}$, $a_{\phi, \mathrm{Q}}$, $b_{\mathrm{M, SF}}$, $b_{\mathrm{M, Q}}$, $b_{\phi, \mathrm{SF}}$ and $b_{\phi, \mathrm{Q}}$.

Throughout this paper, we investigated the sensitivity of the redshift distributions from DES-, HSC- and SDSS-like surveys to these parameters with the goal to identify which galaxy population features need to be well understood for future precision measurements. 
As the result of this and within the context of our forward model, we identified which Schechter parameters are most important to constrain such that the uncertainties of shear measurements are reduced.
Additionally, we explored how the Schechter parameters affected the galaxy observables (magnitudes and colours). In this way, we identified the observables that are most sensitive to the Schechter parameters and that could help to constrain the parameters. 
We refer to section \ref{sec:method} for more details about our methodology.

We found that constraining $a_{\mathrm{M, SF}}$ has the highest priority for simulating the redshift distribution of DES-like surveys. Since upcoming surveys like Euclid and LSST have similar properties as DES, having better constraints of $a_{\mathrm{M, SF}}$ is especially important for future studies that are based on simulated redshift distributions. 
The uncertainty of $a_{\mathrm{M, SF}}$ can affect the mean redshift by up to ${\sim}45$\,\%, which corresponds to an absolute change of ${\sim}0.3$.
The other parameters have less effect on the simulations but the results suggest that $a_\mathrm{M, Q}$, $a_{\phi, \mathrm{SF}}$, $a_{\phi, \mathrm{Q}}$ and $b_\mathrm{M, SF}$ require better constraints as well.

Furthermore, we found that the two-dimensional distributions in colour-magnitude space are generally more sensitive to the parameters than the distributions in colour-colour and magnitude-magnitude space. Therefore, the colour-magnitude distributions can help constraining the parameters, especially $a_\mathrm{M, SF}$, the best.
The results also show that the choice of the colour-magnitude combination has a minor role. 
In the case of a DES-like survey, we found that the $\mathit{(g-r) - i}$, $\mathit{(g-r) - z}$, $\mathit{(g-i) - i}$ and $\mathit{(g-i) - z}$ colour-magnitude combinations are the best options. 
However, HSC-like measurements might be a better choice to constrain the parameters because they have higher sensitivity to the Schechter parameters.

Finally, we will include the error model that is described in section \ref{sec:error-models} to the \verb!skypy.utils! module of the \verb!SkyPy! library \citep{SkyPyCollaboration2020, Amara2021}. 
Additionally, we aim to improve the constraints of the parameters in future projects.

\section*{Acknowledgements}
We would like to thank all our colleagues in the SkyPy Collaboration. We especially acknowledge N. Tessore and R. P. Rollins for many useful discussions in the early stages of this paper. 

The preparation of this manuscript was made possible by a number of software packages: \verb!NumPy! \citep{Harris2020}, \verb!SciPy! \citep{Virtanen2020}, \verb!Astropy! \citep{Price-Whelan2018}, \verb!Matplotlib! \citep{Hunter2007}, \verb!IPython/Jupyter! \citep{Perez2007} and \verb!GetDist! \citep{Lewis2019}.
This research partly developed and made use of \verb!SkyPy!, a Python package for forward modelling astronomical surveys \citep{SkyPyCollaboration2020, Amara2021}.

PS and AA acknowledge support from a Royal Society Wolfson Fellowship grant.

\section*{Data Availability}
The data produced in this work will be available as the corresponding configuration files for the simulations in the \verb!SkyPy! GitHub repository, \url{https://github.com/skypyproject/skypy} and \url{https://doi.org/10.5281/zenodo.4071945}.

The public data used in figures \ref{fig:mUncertaintyDES}, \ref{fig:mUncertaintySDSS} and \ref{fig:mUncertaintyHSC} were obtained from \cite{Drlica-Wagner2018} at \url{https://des.ncsa.illinois.edu/releases/y1a1/gold}, \cite{Ahumada2020} at \url{http://skyserver.sdss.org/CasJobs/} and from \cite{Aihara2019} at \url{https://hsc-release.mtk.nao.ac.jp/datasearch/}, respectively.

\bibliographystyle{mnras}
\bibliography{Galaxy-Forecasting}

\appendix

\section{Plots Redshift Distributions}
\label{sec:appendix_redshift}
\begin{figure*}
    \includegraphics[width=2\columnwidth]{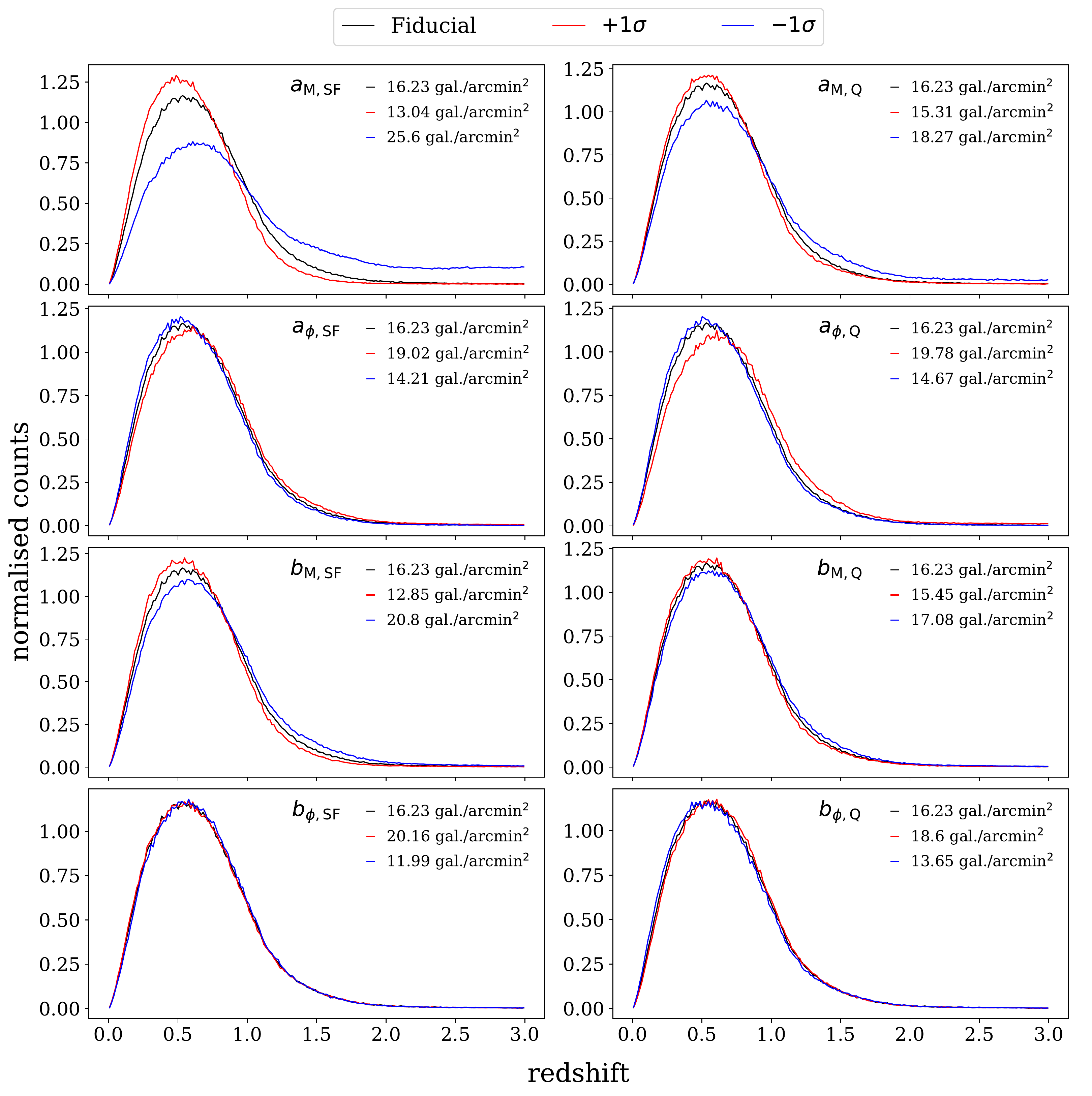}
    \caption{Comparison of the normalised simulated fiducial (black) and test redshift distributions of a \textbf{DES-like survey}. The red and blue lines correspond to increasing and decreasing the Schechter parameter by $1\sigma$, respectively. The number of galaxies per square arcminute is indicated in the legend of each plot. The plots are in agreement with the results in section \ref{sec:redshift_sensitivity}.}
    \label{fig:all_redshift_des}
\end{figure*}

\begin{figure*}
    \includegraphics[width=2\columnwidth]{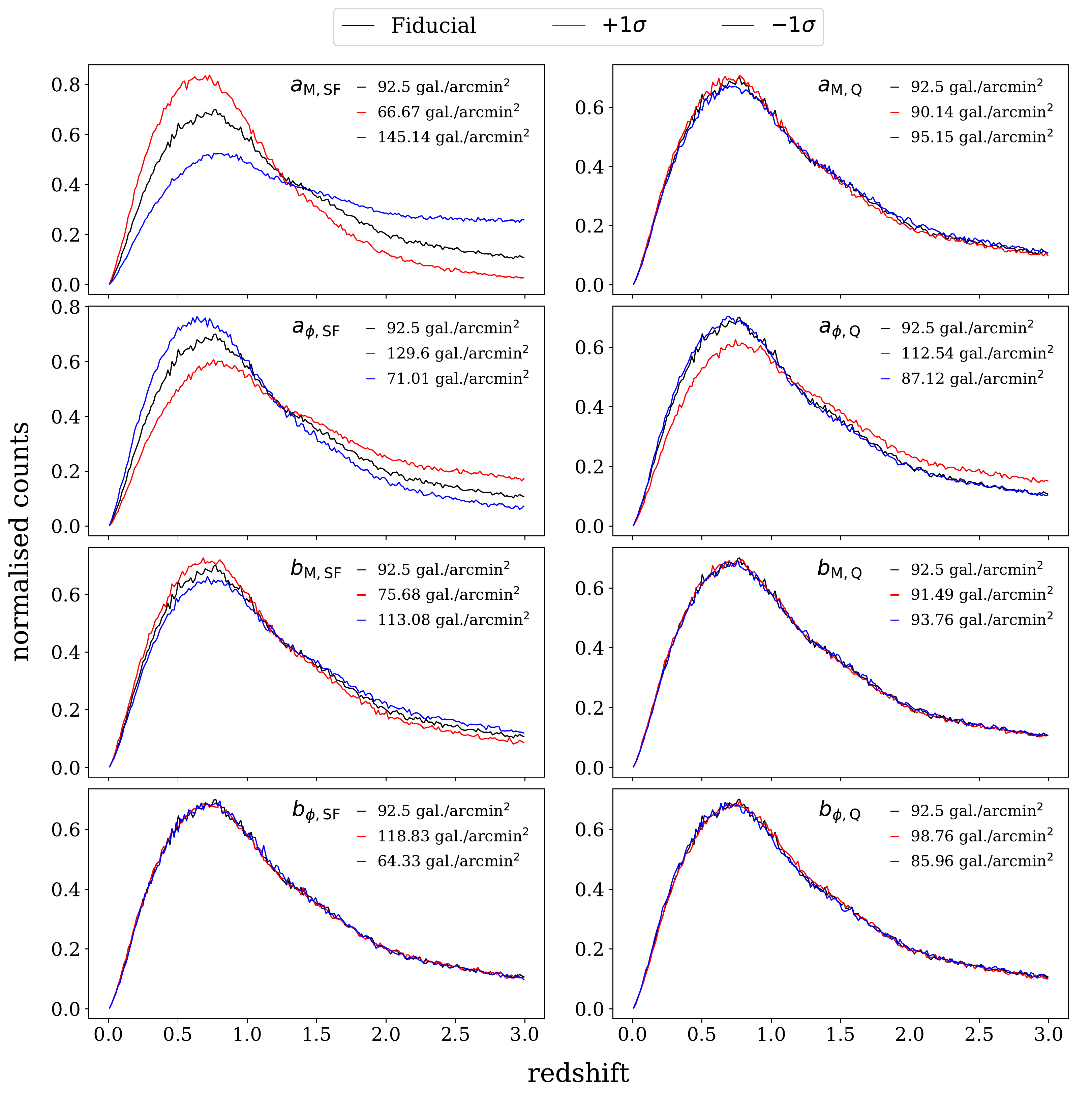}
    \caption{Same as \ref{fig:all_redshift_des} but for an \textbf{HSC-like survey}.}
    \label{fig:all_redshift_hsc}
\end{figure*}

\begin{figure*}
    \includegraphics[width=2\columnwidth]{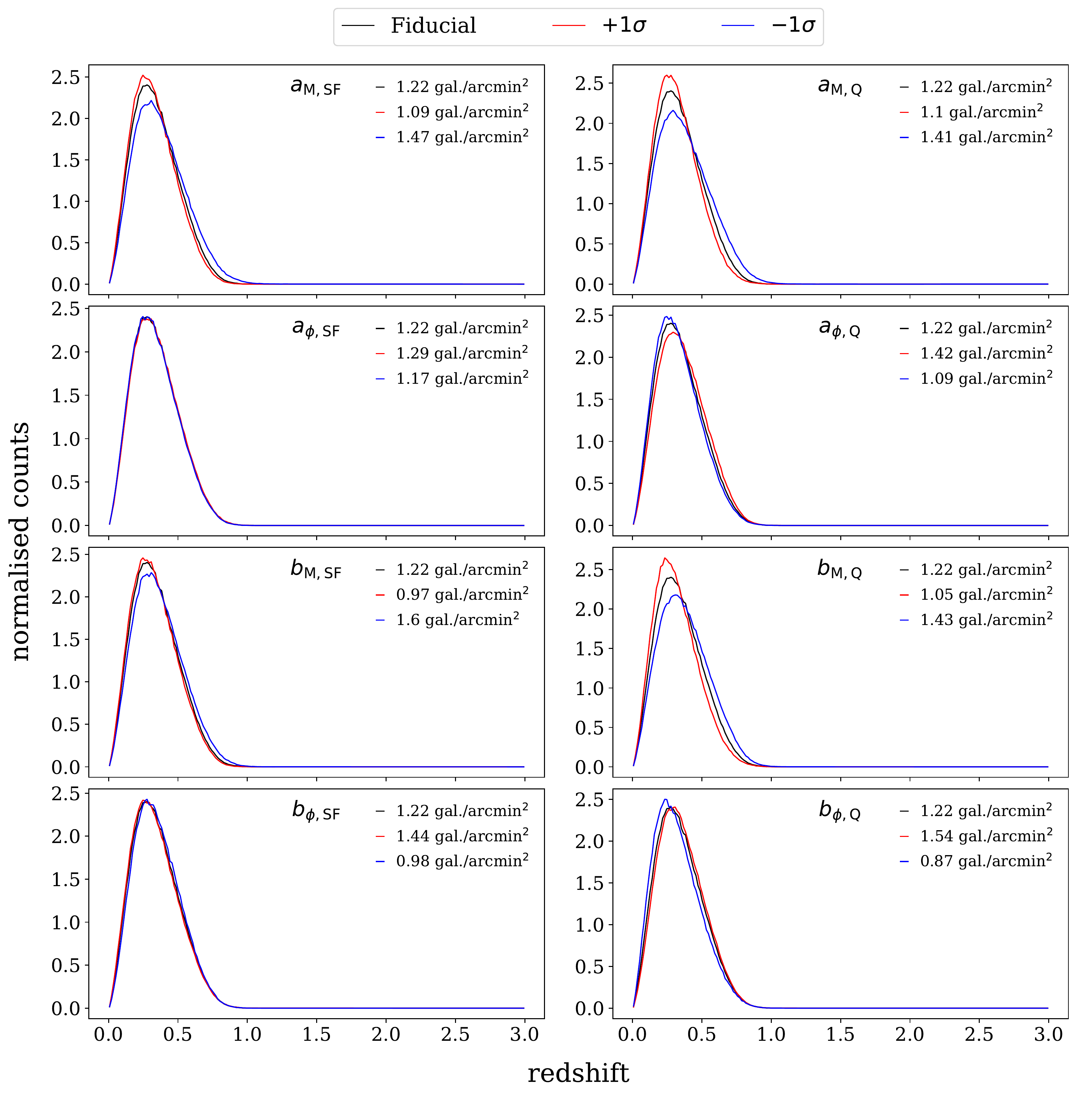}
    \caption{Same as \ref{fig:all_redshift_des} but for an \textbf{SDSS-like survey}.}
    \label{fig:all_redshift_sdss}
\end{figure*}

In this part of the paper, we visually compare the sensitivity of the redshift distributions to all Schechter parameters for the three survey types. 
The individual plots show the fiducial redshift distributions in black. The redshift distributions of the test catalogues after increasing and decreasing the parameters by their given errors are shown in red and blue, respectively.
The legends of each plot contain the galaxy number density for each catalogue.
Figure \ref{fig:all_redshift_des} shows the distributions of a DES-like survey.
Figures \ref{fig:all_redshift_hsc} and \ref{fig:all_redshift_sdss} correspond to an HSC- and an SDSS-like survey, respectively.
The plots are all in agreement with the results in section \ref{sec:redshift_sensitivity}.

\bsp	
\label{lastpage}
\end{document}